\documentclass[a4paper,11pt]{article}
\usepackage{amsfonts}
\usepackage{bbding}
\usepackage{amssymb}
\usepackage{mathrsfs}
\usepackage{float}
\usepackage{amsmath}
\usepackage{amsfonts,amssymb}
\usepackage{graphicx,color}
\usepackage{bm}
\usepackage{mathrsfs}
\catcode`\@=11 \@addtoreset{equation}{section}

\catcode`\@=12

\newtheorem{Lemma}{Lemma}[section]

\newtheorem{Remark}{Remark}[section]

\def\2{{I \hskip -1.0mm I}}
\def\3{{I \hskip -1.0mm I\hskip -1.0mm I}}
\def\4{{I \hskip -0.9mm V}}
\def\6{{V \hskip -1.35mm I}}

\title{Impact-induced tensile waves in a kind of phase-transforming materials}
\author {Shou-Jun Huang\footnote{E-mail address:
sjhuang@mail.ahnu.edu.cn}
\vspace{0.2cm} \\Department of Mathematics, Anhui Normal University\\
 Wuhu 241000, China }
\date{ }
\begin{document}
\maketitle
\begin{abstract}
This paper concerns the global propagation
 of impact-induced tensile waves in  a kind of phase-transforming materials.
 It is well-known that the governing system of partial differential equations
 is hyperbolic-elliptic and the initial-boundary value problem is not well-posed
 at all levels of loading.  By making use of fully nonlinear stress-strain curve to model this material,
  Dai and Kong succeeded in constructing a physical solution of the above initial-boundary value problem. For
  the impact of intermediate range, they assumed that $\beta<3\alpha$ in the stress-response function for simplicity.
   In this paper, we  revisit the impact problem and consider the propagation of
   impact-induced tensile waves for all values of the parameters $\alpha$ and $\beta$. The physical solutions for all levels of loading are obtained completely.
\vskip 6mm


\vskip 3mm

\end{abstract}
\newpage
\baselineskip=7mm

\section{Introduction}
For a thin bar or rod made of a kind of phase-transforming
materials, and placed in an equilibrium state of uniaxial tension,
the curve of the nominal stress and longitudinal strain is concave
from small to moderate strains  but strongly convex for large
strains. This kind of phase-transforming materials have many
applications and have been studied by many people (e.g.,
\cite{rkj}-\cite{dk1} and references therein). It is well-known that
the system of conservation laws governing the dynamics of the
elastic bar is hyperbolic for a standard material for which the
nominal stress is a monotonically increasing function of the
longitudinal strain, and is hyperbolic-elliptic for a typical phase
transforming material. Abeyaratne et al. \cite{rkj} considered the
impact-induced phase transition problem in a semi-infinite bar with
a given velocity $-V$ at the end. By using tri-linear approximation
for the stress-strain curve, they constructed a two-wave solution
involving a shock wave and a phase boundary. Knowles \cite{jk}
studied the impact-induced tensile waves in a one-dimensional
semi-infinite bar made of a rubberlike material. Knowles showed that
there are three regimes of response related to the loading intensity
and for the impact of intermediate strength, under the hypotheses
such as maximally dissipative kinetics or dissipation free kinetics,
he selected the unique admissible solution among one-parameter
family of solutions.

Recently, Dai and Kong \cite{dk} used a fully nonlinear
stress-strain curve instead of a linearized tri-linear curve in
literature, and with the aid of the uniqueness condition on phase
boundary derived by Dai \cite{dai}, they succeeded in constructing
the unique physical solution of the initial- boundary value problem
for the loading at all levels. In particular, for the intermediate
impact velocity,
 Dai and Kong \cite{dk} assumed that the two parameters in the stress-response function (see (\ref{3}) below)
 satisfy $\beta<3\alpha$, and based on this, they established the unique physical solution containing
  a centered rarefaction wave followed by a phase-boundary. They also pointed out that the characteristic speeds
in the front- and back-states of the phase boundary are equal.

In this paper, we revisit the initial-boundary value problem
mentioned above for a kind of phase transforming materials.  The
inequality $\beta<3\alpha$ will be refined  and replaced by another
one and then all physically admissible solutions for all levels of
loading are obtained completely. Especially, for the intermediate
impact, we also construct a physical solution involving a centered
rarefaction wave followed by a phase boundary. Comparably, the
geometric character of this phase boundary exhibits a different
manner.

The paper is organized as follows. In Section 2, we recall the initial-boundary value problem modeling the impact-induced tensile waves in a
semi-infinite bar made of a kind of phase-transforming material.
 Section 3 is devoted to the establishment of admissible solutions for all
  loading at all levels.
  Particularly, in the intermediate case, we completely construct
  physical solutions for all parameters in the stress-response function.
We conclude the paper in section 4 and  some related remarks are
also given.

\section{The initial-boundary value problem}

Consider the time-dependent deformation of a one-dimensional elastic bar from the natural state which occupies the nonnegative $x$-axis.
At the time $t$, a particle at $x$ in the reference state is carried out to $x+u(t,x)$, where $u$ denotes the  longitudinal displacement of
the particle. Hence the strain $\gamma(t,x)$ and the particle velocity $v(t,x)$ are defined by $\gamma=u_x$ and $v=u_t$ respectively.
To ensure the deformation is one-to-one,  we suppose $\gamma>-1$. Let $\sigma(t,x)$ stands for the nominal stress at the time $t$ for this particle. In Lagrangian description, the governing system of equations of motion is given by
\begin{equation}\label{1}\rho v_t=\sigma_x,\quad \gamma_t=v_x,\end{equation}
where $\gamma,v$ are smooth, and the jump conditions
\begin{equation}\label{2}\dot{s}[\gamma]+[v]=0,\quad \dot{s}\rho[v]+[\sigma]=0\end{equation}
at a moving strain discontinuity whose referential location is $x=s(t)$ at time $t$., in which
$[f]=f(t,s(t)+0)-f(t,s(t)-0)$, $\rho$ is the constant density and $\dot{s}=s'(t)$ is denoted by the Lagrangian velocity of the discontinuity.

The material of the bar is taken to be elastic so that $\sigma=\sigma(\gamma)$, where the stress-response relation $\sigma=\sigma(\gamma)$ is assumed to be a $C^2$-smooth function. Thus, the system (\ref{1}) is hyperbolic-elliptic for typical phase-transforming materials for which  the stress-strain curve generally has a peak-valley combination.

 As in Dai and Kong \cite{dk}, this paper concerns the following nonlinear stress-strain relation in order to model two phase materials
\begin{equation}\label{3}\sigma(\gamma)=E\left(\frac{\gamma^3}{3}-\frac12(\alpha+\beta)\gamma^2+\alpha\beta\gamma\right), \end{equation}
where $E$ is the Young's modulus for infinitesimal strains, $\alpha$ and $\beta$ are constants satisfying
 \begin{equation}\label{4}0<\alpha<\beta<\infty.\end{equation}
In general, some parts of the curve may lie below the strain axis.
The restrictions of the stress-response function (\ref{3}) to the
intervals $[0,\alpha]$ and $[\beta,\infty)$ are called
$\alpha$-branch and $\beta$-branch respectively. Obviously,
$\sigma(\gamma)$ is concave for $-1<\gamma<(\alpha+\beta)/2$ and
convex for $\gamma>(\alpha+\beta)/2$.  see Fig.1.

\vspace{5mm}
\begin{figure}\begin{center}
\includegraphics{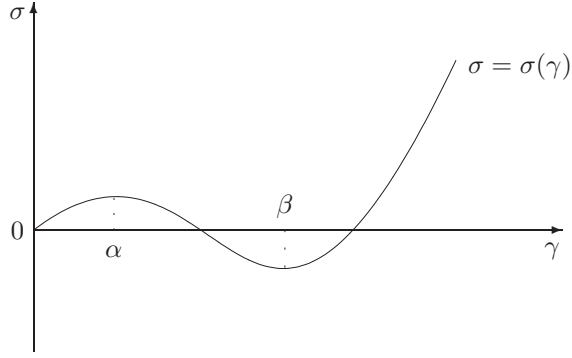}
 \renewcommand{\figurename}{Fig.}\caption{\it{The stress-strain curve $\sigma(\gamma)$}}
\end{center}
\end{figure}

The sound-wave speed $c(\gamma)$ for the bar characterized by (\ref{3}) is given by
\begin{equation}\label{5}c(\gamma)=\sqrt{\sigma'(\gamma)/\rho}=c_0\sqrt{(\gamma-\alpha)(\gamma-\beta)/\alpha\beta}\quad\text{for}
\;\gamma\in[0,\alpha]\cup[\beta,\infty),\end{equation}
where $c_0$ is the speed of small amplitude waves at the undeformed state
$$c_0=c(0)=\sqrt{E\alpha\beta/\rho}.$$

 As in \cite{dk},  for the system (\ref{1}), the impact problem corresponds to the following initial-boundary conditions
\begin{equation}\label{6}\gamma(0,x)=0,\;v(0,x)=0\quad \text{for} \;x>0,\end{equation}
\begin{equation}\label{7}v(t,0)=-V\quad \text{for} \;t>0,\end{equation}
where $V$ is the impact velocity and assumed throughout to be positive.
We shall seek the solution of the initial boundary value problem (\ref{1}), (\ref{6})-(\ref{7}) on the first quadrant.

\section{The impact-induced tensile waves}
In this section, we exhibit the structure of solutions of the impact problem (\ref{1}), (\ref{6})-(\ref{7}) for all values of $\alpha, \beta$ in the stress-response function (\ref{3}) under the assumption (\ref{4}).

Introduce
\begin{equation}\label{8}V_*=\int_0^{\alpha}c(\gamma)d\gamma=c_1\int_1^h\sqrt{\zeta^2-1}d\zeta,
\;V_{**}=2c_1h\sqrt{\frac{h^2}{3}-1},\;V_{***}=3c_1h\sqrt{h^2-1},\end{equation}
where
\begin{equation}\label{9}c_1\triangleq\frac{(\beta-\alpha)^2}{4\sqrt{\alpha\beta}}c_0,
\quad h\triangleq\frac{\beta+\alpha}{\beta-\alpha}.\end{equation}
\begin{Remark}The values of constants $V_*, V_{**}$ and $V_{***}$ coincide with those in \cite{dk}.\end{Remark}
 It is easy to derive the following lemma.
\begin{Lemma}Under the assumption (\ref{4}), $V_*, V_{***}$ are positive constants and satisfy
\begin{equation}\label{n1}0<V_{*}<V_{***}<\infty;\end{equation}
if furthermore we assume  that
 \begin{equation}\label{n2}\beta<\frac{h_*+1}{h_*-1}\alpha,\end{equation}
where $h_*\in(\sqrt{3},2)$ is the unique solution to $V_*=V_{**}$,
then $V_{**}$ makes sense and it holds that
\begin{equation}\label{n3}0<V_{*}<V_{**}<V_{***}<\infty.\end{equation} \end{Lemma}

Although for weak impact $(V\leqslant V_*)$ and strong impact $V\geqslant V_{***}$,
 the solutions for the impact problem are the same as constructed in Dai and Kong \cite{dk},
   we still state them in the sequel for completeness.

\subsection{Weak impacts}
As in \cite{dk}, the solution contains a centered rarefaction wave and is given by

\begin{equation}\label{10}(v(t,x),\gamma(t,x))=\left\{ \begin{aligned}& (-V,\gamma_1)\quad  &&\text{for}\; 0\leqslant \xi\leqslant\xi_1,
\\&(\hat{v}(\xi),\hat{\gamma}(\xi))\quad  &&\text{for}\; \xi_1\leqslant \xi\leqslant\xi_2, \\&
(0,0)\quad  &&\text{for}\; \xi\geqslant\xi_2,\end{aligned} \right.\end{equation}
where $\xi=x/t,\,\xi_2=c_0$ and $\hat{v},\, \hat{\gamma}, \,\gamma_1, \,\xi_1$ are defined by
\begin{equation}\label{11}c(\hat{\gamma}(\xi))=\xi,\;\hat{v}(\xi)=\int_{\xi}^{c_0}\zeta\hat{\gamma}'(\zeta)d\zeta,\;
\int_0^{\gamma_1}c(\gamma)d\gamma=V\;\text{and}\;\gamma_1=\hat{\gamma}(\xi_1).\end{equation}
The interval $(0,V_{*}]$ is the regime of weak impacts. As the impact velocity $V\in(0,V_{*}]$, the strain $\gamma_1$ satisfies $\gamma_1\leqslant\alpha$.

\subsection{Strong impact}
As in \cite{dk}, the solution contains a shock wave and is given by
\begin{equation}\label{12}(v(t,x),\gamma(t,x))=\left\{ \begin{aligned}& (-V,\gamma^-)\quad  &&\text{for}\; 0\leqslant x<\dot{s}t,
 \\&(0,0)\quad  &&\text{for}\; x>\dot{s}t,\end{aligned} \right.\end{equation}
where $\gamma^-$ and $\dot{s}$ are defined by
\begin{equation}\label{13}\gamma^-c_0\sqrt{{\gamma^-}^2/3-(\alpha+\beta)\gamma^-/2+\alpha\beta}=V\sqrt{\alpha\beta}
\qquad \text{and}\qquad \dot{s}=\frac{V}{\gamma^-}.\end{equation}
Obviously, this pure-shock-wave solutions are constructed for which $\gamma^->\frac32(\alpha+\beta)$ and therefore $V>V_{***}$. Moreover, the shock wave is supersonic in the Lagrangian sense with respect to the undisturbed state ahead of it, i.e. $c_0<\dot{s}<c(\gamma^-)$. In particular,
when $\gamma^-=\frac32(\alpha+\beta)$, the impact velocity $V=V_{***}$ and the discontinuity $x=\dot{s}t$ is called a degenerate shock wave since $\dot{s}=c_0$ and $\dot{s}<c(\gamma^-)$.

\subsection{Impact of intermediate strength}
In this subsection, we turn to construct the solution of the initial-boundary value problem (\ref{1}), (\ref{6})-(\ref{7})  to fill the gap of impact velocity $V_*<V< V_{***}$.
We now attempt to construct one solution involving  a centered rarefaction wave followed by a phase-boundary, as shown in Fig. 2. Precisely speaking, we propose the following form of solution
\begin{equation}\label{14}(v(t,x),\gamma(t,x))=\left\{ \begin{aligned}& (-V,\gamma^-)\quad  &&\text{for}\; 0\leqslant x<\dot{s}t,
\\&(v^+,\gamma^+)\quad &&\text{for}\;\dot{s}t<x\leqslant \xi_1t,
\\&(\hat{v}(\xi),\hat{\gamma}(\xi))\quad  &&\text{for}\; \xi_1t\leqslant x\leqslant\xi_2t, \\&
(0,0)\quad  &&\text{for}\; x\geqslant\xi_2t,\end{aligned} \right.\end{equation}
where $\gamma^-, \,\dot{s}, \,\xi_1,\, \hat{v}$ and $\hat{\gamma}$ are to be determined and, by continuity, $\xi_2=c_0, \gamma^+=\hat{\gamma}(\xi_1)$
and $v^+=\hat{v}(\xi_1)$.

\begin{figure}[th!]\begin{center}
\includegraphics{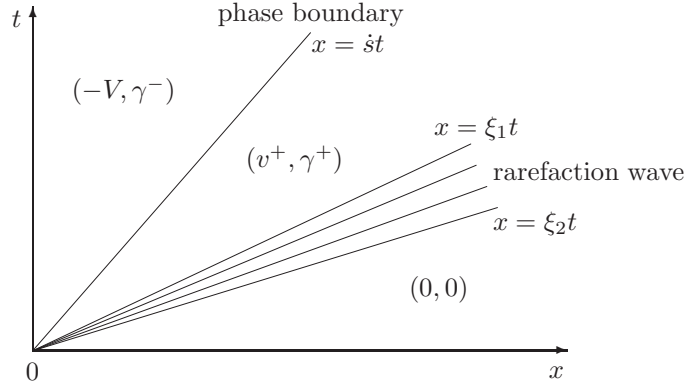}
 \renewcommand{\figurename}{Fig.}\caption{\it{Solution with a centered rarefaction wave followed by a phase boundary}}
\end{center}
\end{figure}

Assuming that $\gamma$ and $v$ are continuous everywhere except across the phase boundary, we  have $\xi_1=c(\gamma^+)$.
As in the discussion  of the centered rarefaction wave arising from weak impact \cite{dk}, we obtain that
\begin{equation}\label{15}v^+=\hat{v}(\xi_1)=\int^{c_0}_{\xi_1}\zeta\hat{\gamma}'(\zeta)d\zeta=-\int_0^{\gamma^+}c(\gamma)d\gamma.\end{equation}

Since the phase boundary is to trail the centered rarefaction wave, we require that
 \begin{equation}\label{16}0<\dot{s}\leqslant\xi_1=c(\gamma^+).\end{equation}
Using (\ref{15}) in the jump conditions (\ref{2}) applied at the phase boundary $x=\dot{s} t$ leads to
\begin{equation}\label{17}(\gamma^+-\gamma^-)\dot{s}-\left\{\int_0^{\gamma^+}c(\gamma)d\gamma-V\right\}=0.\end{equation}
\begin{equation}\label{18}\sigma(\gamma^+)-\sigma(\gamma^-)-\rho\dot{s}\left\{\int_0^{\gamma^+}c(\gamma)d\gamma-V\right\}=0\end{equation}
These are two equations for the front- and back-state strains $\gamma^{\pm}$ at the phase boundary; they involve the phase boundary speed $\dot{s}$
as an unknown function of $V$, while the impact velocity $V$ as a given datum.

Eliminating the common contents of the braces between (\ref{17}) and (\ref{18}) yields the standard formula relating the phase boundary speed and the slope of the chord connecting the two points on the stress-strain curve that correspond to the front- and back-states of the phase boundary
\begin{equation}\label{19}\rho\dot{s}^2=\frac{\sigma(\gamma^+)-\sigma(\gamma^-)}{\gamma^+-\gamma^-}.\end{equation}
It is easy to see that the equations (\ref{17})-(\ref{18}) are equivalent to either (\ref{17}) or (\ref{18}) together with (\ref{19}).

For the material governed by (\ref{3}), (\ref{17}) and (\ref{19}) take the respective forms
\begin{equation}\label{20}(\gamma^+-\gamma^-)\dot{s}+\int_{\gamma^+}^{\alpha}c(\gamma)d\gamma+V-V_*=0,\end{equation}
\begin{equation}\label{21}\begin{aligned}&\left(\gamma^+-\frac{\alpha+\beta}{2}\right)^2
+\left(\gamma^+-\frac{\alpha+\beta}{2}\right)\left(\gamma^--\frac{\alpha+\beta}{2}\right)
\\&+\left(\gamma^--\frac{\alpha+\beta}{2}\right)^2=\frac{3\alpha\beta}{c_0^2}\dot{s}^2+3\left(\frac{\beta-\alpha}{2}\right)^2.\end{aligned}\end{equation}
We shall view (\ref{20}) and (\ref{21}) as the system (\ref{17}) and
(\ref{18}).

For each $V$ between $V_*$ and $V_{***}$, we shall show that there is a one-parameter family of solutions to the initial-boundary value problem (\ref{1}),
(\ref{6})-(\ref{7}). So a mechanism is needed to select one physical unique solution.
Abeyaratne and Knowles introduced the concept of driving force, which is defined via the dissipation rate (cf. \cite{rj}-\cite{rj2}), and the kinetic relations to select the unique admissible solution.
The dissipation rate can be written as
\begin{equation}\label{22}D(t)=f(t)s'(t),\end{equation}
where $f(t)$ is called the driving force per unit cross-sectional area acting at time $t$ on the moving strain discontinuity. For a phase-transforming material governed by
 (\ref{3}), direct computation shows that $f$ is given by
\begin{equation}\label{23}f=\frac{E}{12}(\gamma^+-\gamma^-)^3\left\{(\alpha+\beta)-(\gamma^++\gamma^-)\right\}.\end{equation}
Since $\dot{s}>0$, the physical admissibility  in the sense of Knowles (see \cite{jk} for definition) requires that $D(t)\geqslant0$. Since in (\ref{23}) one has $\gamma^->\gamma^+$, it implies that
\begin{equation}\label{24}\gamma^++\gamma^-\geqslant\alpha+\beta.\end{equation}
 If in (\ref{16}) one represents $\dot{s}$ through (\ref{21}) and $c(\gamma^+)$ through (\ref{5}), one finds, after simplification, the further restriction
\begin{equation}\label{25}2\gamma^++\gamma^-\leqslant\frac32(\alpha+\beta).\end{equation}
Moreover, in order to keep $\dot{s}>0$ in (\ref{21}), direct computation gives the following  restriction
\begin{equation}\label{26}\gamma^-\geqslant g(\gamma^+)\triangleq\frac{\alpha+\beta}{2}+\frac12\left[\frac{\alpha+\beta}{2}-\gamma^++
\sqrt{3\left(\frac{3\beta-\alpha}{2}-\gamma^+\right)\left(\frac{\beta-3\alpha}{2}+\gamma^+\right)}\,\right]\end{equation}
Pairs $(\gamma^+,\gamma^-)$ satisfying the inequalities (\ref{24})-(\ref{26}) as well as $\gamma^+\leqslant\alpha, \gamma^-\geqslant\beta$ correspond to points in the quadrilateral BCEF, the triangles BCE and BCF in the $\gamma^+,\gamma^-$-plane  for three cases respectively, see Fig 3.
\vspace{5mm}
\begin{figure}[!ht]
\centering
\includegraphics[width=0.33\textwidth]{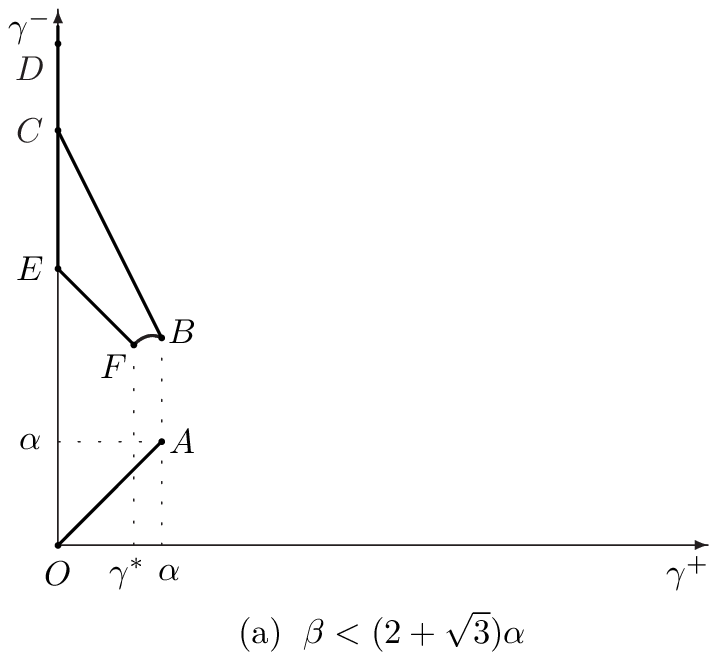}\includegraphics[width=0.33\textwidth]{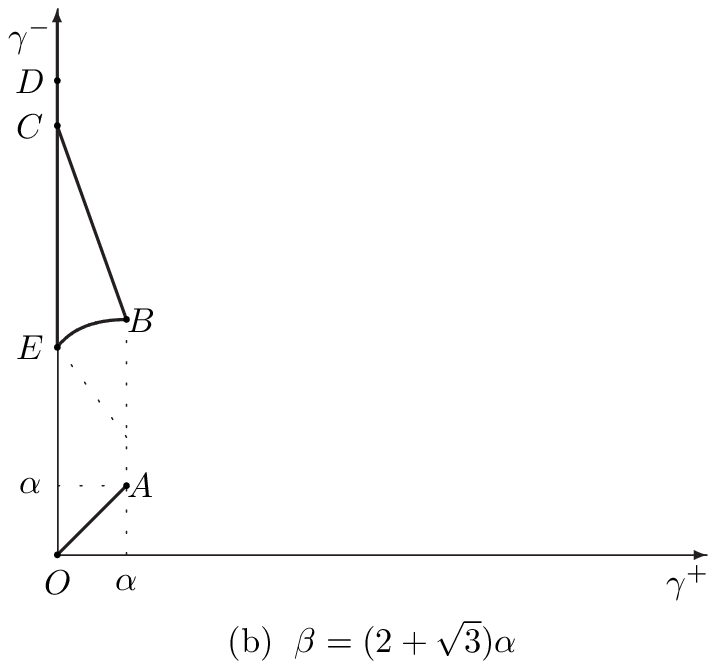}\includegraphics[width=0.33\textwidth]{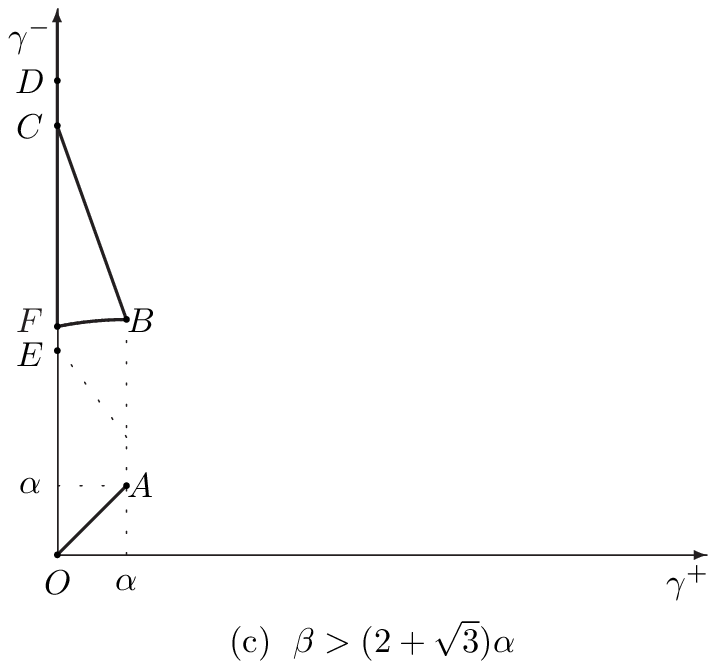}
\renewcommand{\figurename}{Fig.}\caption{\small\it{ Pairs $(\gamma^+,\gamma^-)$ satisfying (\ref{17})-(\ref{18}) must respectively
correspond to points in the quadrilateral BCEF, the triangles BCE and BCF in the $\gamma^+,\gamma^-$-plane for the cases (a), (b) and (c). Points of the form $(0,\gamma^-)$ with $\gamma^-\geqslant\frac32(\alpha+\beta)$, lying on the vertical axis above the point C, correspond to pure shock wave solution. Points on the line OA: $\gamma^+=\gamma^-$ with $0<\gamma^+<\alpha$ correspond to the centered rarefaction wave. The curve BE or BF is given by $\gamma^-=g(\gamma^+)$.}}\end{figure}

The quadrilateral and triangular regions in Fig.3, along with the portions of straight lines that represent the centered rarefaction wave and pure shock wave solutions, can be mapped through (\ref{20}) and (\ref{21}) into a plane in which the natural Cartesian coordinates are the dimensionless phase boundary velocity $\dot{s}/c_2$ and the dimensionless impact velocity $V/c_1$, where $c_2=2c_1/(\beta-\alpha)$.
Consider first the centered rarefaction wave solutions. Setting $\gamma^+=\gamma^-=\frac{\alpha+\beta}{2}-\frac{\beta-\alpha}{2}\eta$ in (\ref{20}) and (\ref{21}) gives rise to the parametric equations
\begin{equation}\label{27}\left\{\begin{aligned}\frac{\dot{s}}{c_2}&=\sqrt{\eta^2-1},\\
\frac{V}{c_1}&=\int_{\eta}^h\sqrt{\zeta^2-1}\,d\zeta,\end{aligned}\right.\;1\leqslant\eta\leqslant h.\end{equation}

Next we map the segment BC  for which $\gamma^-=\frac32(\alpha+\beta)-2\gamma^+$ with $0<\gamma^+\leqslant\alpha$.
By (\ref{20})-(\ref{21}), the image of this segment in the dimensionless $\dot{s}, V$-plane is the following curve
 \begin{equation}\label{28}\left\{\begin{aligned}\frac{\dot{s}}{c_2}&=\sqrt{\eta^2-1},\\
\frac{V}{c_1}&=3\eta\sqrt{\eta^2-1}+\int_{\eta}^h\sqrt{\zeta^2-1}\,d\zeta,\end{aligned}\right.\;1\leqslant\eta\leqslant h.\end{equation}
From (\ref{20})-(\ref{21}), the image of segment CD for which $\gamma^+=0, \gamma^-\geqslant\frac32(\alpha+\beta)$ is found to be the
curve represented by
 \begin{equation}\label{29}\left\{\begin{aligned}\frac{\dot{s}}{c_2}&=\sqrt{\frac13(\varphi^2-h\varphi+h^2)-1},\\
\frac{V}{c_1}&=(\varphi+h)\frac{\dot{s}}{c_2},\end{aligned}\right.\;\;\;\varphi\geqslant 2h,\end{equation}
where \begin{equation}\label{30}\varphi=\frac{2\gamma^--(\alpha+\beta)}{\beta-\alpha}.\end{equation}
By the same method, the segments CE in the Fig 3. (a) and (b) can be mapped into the following parameterized curve
 \begin{equation}\label{31}\left\{\begin{aligned}\frac{\dot{s}}{c_2}&=\sqrt{\frac13(\varphi^2-h\varphi+h^2)-1},\\
\frac{V}{c_1}&=(\varphi+h)\frac{\dot{s}}{c_2},\end{aligned}\right.\;\;\;h\leqslant\varphi\leqslant 2h.\end{equation}
Form (\ref{20}), (\ref{21}) and (\ref{26}), the segment CF in Fig 3. (c) may now be mapped into the curve in the dimensionless $\dot{s}, V$-plane
 \begin{equation}\label{32}\left\{\begin{aligned}\frac{\dot{s}}{c_2}&=\sqrt{\frac13(\varphi^2-h\varphi+h^2)-1},\\
\frac{V}{c_1}&=(\varphi+h)\frac{\dot{s}}{c_2},\end{aligned}\right.\;\;\;\frac12[h+\sqrt{3(4-h^2)}]\leqslant\varphi\leqslant 2h.\end{equation}
In Fig 3. (a), the image of segment EF now takes the following form
  \begin{equation}\label{33}\left\{\begin{aligned}\frac{\dot{s}}{c_2}&=\sqrt{\frac{\eta^2}{3}-1},\\
\frac{V}{c_1}&=2\eta\sqrt{\frac{\eta^2}{3}-1}+\int_{\eta}^h\sqrt{\zeta^2-1}\,d\zeta,\end{aligned}\right.\;\sqrt{3}\leqslant\eta\leqslant h.\end{equation}
Finally, the curve BF or BE in Fig 3. can be represented in the following parametric form
 \begin{equation}\label{34}\left\{\begin{aligned}\frac{\dot{s}}{c_2}&\equiv0,\\
\frac{V}{c_1}&=\int_{\eta}^h\sqrt{\zeta^2-1}\,d\zeta.\end{aligned}\right.\end{equation}
The  curves BF and BE in Fig 3. (a)-(b), BF in Fig 3. (c) are parameterized through (\ref{34}) with $\eta\in[1,\sqrt{3}]$ and $[1,h]$ respectively.

The following Fig. 4 and Fig. 5 show the image just constructed in the dimensionless $\dot{s}, V$-plane of the locus in the $\gamma^+, \gamma^-$-plane describing the various types of solutions as presented in Fig 3.

\begin{figure}[!ht]
\centering
\includegraphics[width=0.33\textwidth]{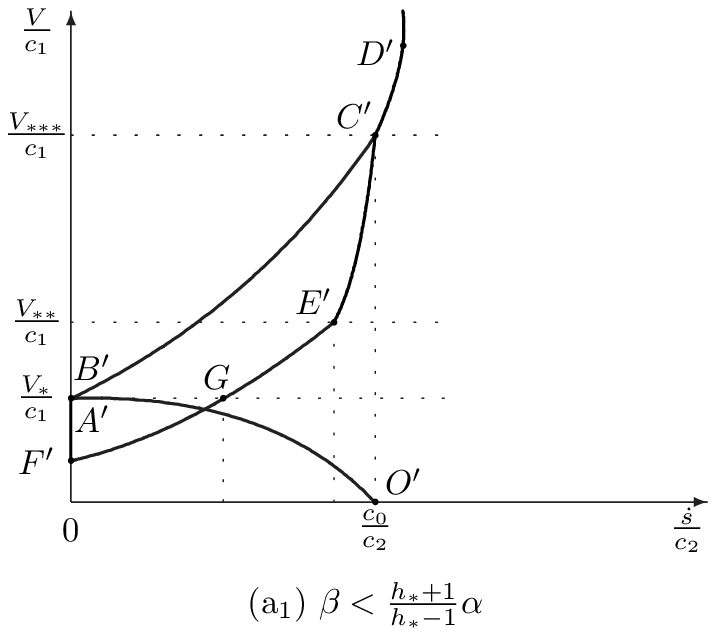}\includegraphics[width=0.33\textwidth]{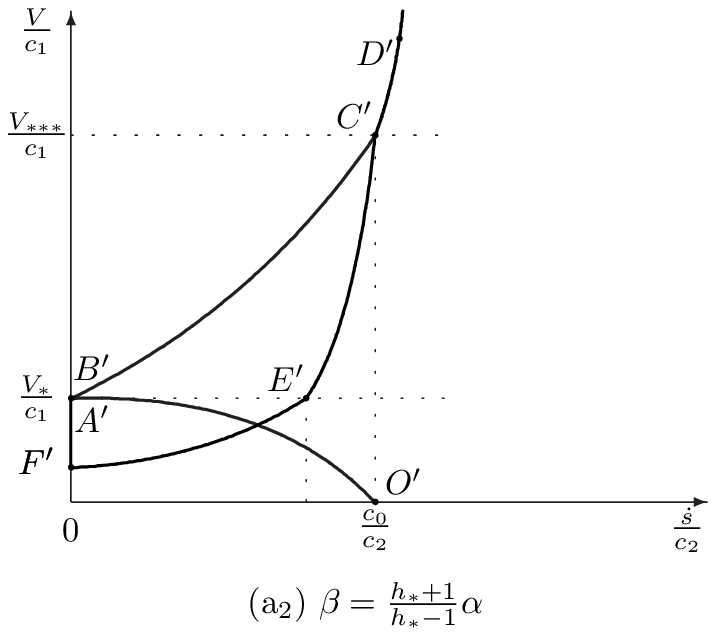}\includegraphics[width=0.33\textwidth]{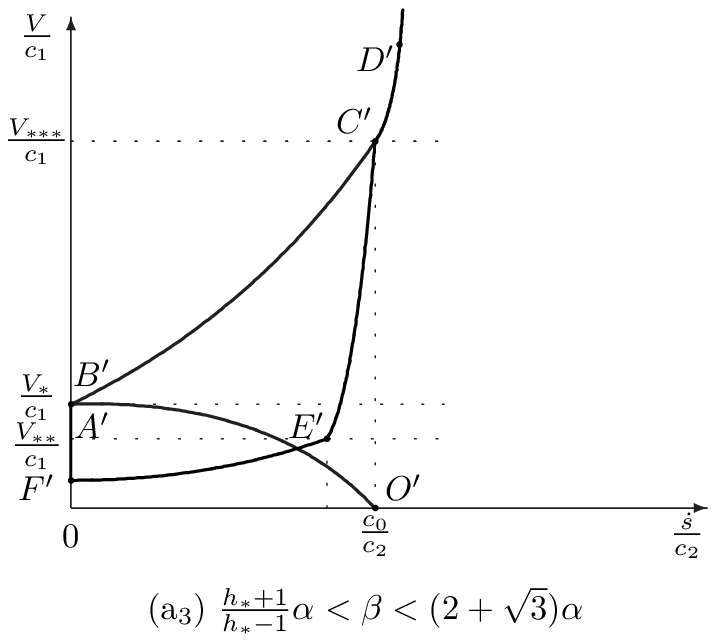}
\renewcommand{\figurename}{Fig.}\caption{\small\it{ Pairs $(\dot{s}/c_2,V/c_1)$ corresponding to points $(\gamma^+,\gamma^-)$ in the locus of Fig. 3 (a). Cases (a$_1$), (a$_2$) and (a$_3$) correspond to $V_*<V_{**}$, $V_*=V_{**}$ and $V_*>V_{**}$ respectively.}}\end{figure}

\begin{figure}[!ht]
\centering
\includegraphics[width=0.4\textwidth]{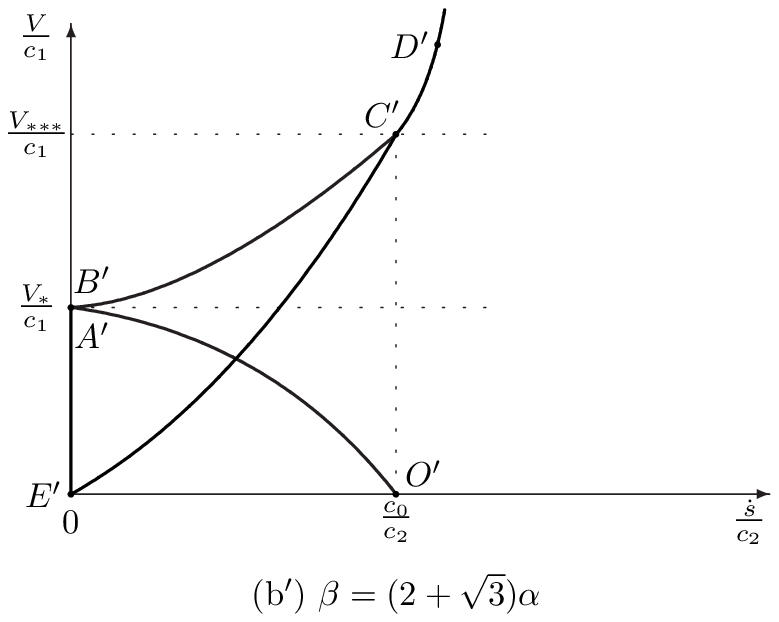}\quad\includegraphics[width=0.4\textwidth]{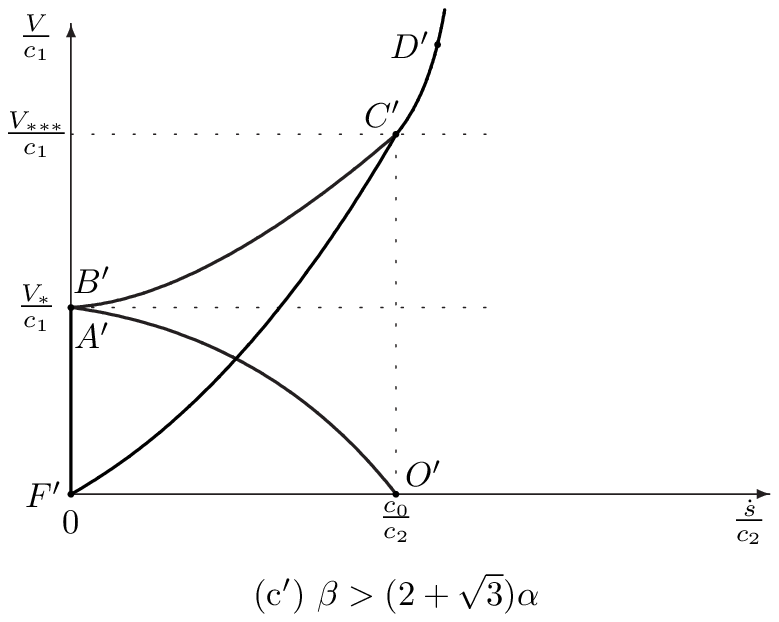}
\renewcommand{\figurename}{Fig.}\caption{\small\it{Pairs $(\dot{s}/c_2,V/c_1)$ corresponding to points $(\gamma^+,\gamma^-)$ in the locus of Fig. 3 (b) and (c). In the present cases, the impact velocity $V_{**}$ does not make sense.}}\end{figure}

It is worthy pointing out that if, for given $\dot{s}$ and $V$, there is a physically admissible solution $\gamma^+, \gamma^-$ of the equations
(\ref{20}), (\ref{21}) with $0\leqslant\gamma^+\leqslant\alpha$ and for which $0<\dot{s}\leqslant c(\gamma^+)$, then it has been shown that
the point corresponding to $\dot{s}, V$ must lie in the  quadrilaterals B$'$C$'$E$'$F$'$ or the triangular regions B$'$C$'$E$'$, B$'$C$'$F$'$ in Fig. 4 and Fig. 5. While, the converse is not proved. This means that for every pair $\dot{s}, V$ corresponding to a point in the above regions, there is
an acceptable solution $\gamma^+, \gamma^-$ of (\ref{20}), (\ref{21}).

Now we  refine the inequality $\beta<3\alpha$ of Dai and Kong \cite{dk} into (\ref{n2}). In other words, under the assumption $\beta<\frac{h_*+1}{h_*-1}\alpha$,
by means of the same method, we can also construct the unique physical solution involving a centered rarefaction wave followed by a phase boundary, which moreover satisfies the following unique condition on the phase boundary derived by Dai \cite{dai}
\begin{equation}\label{35}\gamma^++\gamma^-=\alpha+\beta.\end{equation}
This solution corresponds to the arc GE$'$ in Fig 4. (a$_1$), while
the curve O$'$A$'$ corresponds to the centered rarefaction wave for
weak impact; the curves C$'$E$'$ and C$'$D$'$ relate to the
admissible solution and pure shock wave solution presented in
\cite{dk}. Here we would like to point out that the mixed centered
rarefaction wave and phase boundary solution presented in \cite{dk}
in fact satisfies the dissipation-free kinetics introduced by
Abeyaratne and Knowles (cf. \cite{rj}-\cite{rj2} or \cite{jk}),
i.e., the driving force $f\equiv0$.

In the reminder of this subsection, we aim to construct the unique
physical solution to (\ref{1}), (\ref{6}) and (\ref{7}) for the left
case $\beta\geqslant\frac{h_*+1}{h_*-1}\alpha$. For the impact of
intermediate strength, there is also a one-parameter family of
two-wave solutions (see Fig. 4 and Fig .5). In order to select the
unique admissible solution, we apply the maximally dissipative
kinetics, which implies that the phase boundary speed should satisfy
(cf. \cite{jk})
\begin{equation}\label{36}\dot{s}=c(\gamma^+).\end{equation}
The equation (\ref{36}) means that the phase boundary in fact coincides with the tail of the centered rarefaction wave.
By (\ref{19}) and (\ref{36}), we have the following important identity
\begin{equation}\label{37}2\gamma^++\gamma^-=\frac32(\alpha+\beta).\end{equation}

Next we shall show that the equations (\ref{19}) and (\ref{20}) have a solution $(\gamma^+,\gamma^-)$ satisfying
$\gamma^+\in (0,\alpha).$
Making use of (\ref{36}) and (\ref{37}), it follows from (\ref{20}) that
\begin{equation}\label{38}\left[\frac32(\alpha+\beta)-3\gamma^+\right]c(\gamma^+)=\int_{\gamma^+}^{\alpha}
c(\gamma)d\gamma+V-V_*.\end{equation}
Let $$G(\gamma^+)=\left[\frac32(\alpha+\beta)-3\gamma^+\right]c(\gamma^+)-\int_{\gamma^+}^{\alpha}
c(\gamma)d\gamma+V_*-V,$$
it is easy to observe that $G(\gamma^+)$ is a monotonically decreasing function as $\gamma^+\in(0,\alpha)$.
Moreover, we have
$$G(\alpha)=V_*-V<0,\quad G(0)=\frac32(\alpha+\beta)c_0-V=V_{***}-V>0.$$
 Thus, it is found that there is a unique $\gamma^+\in (0,\alpha)$ such that  (\ref{38}) holds.
 Once $\gamma^+$ has been solved, $\gamma^-$ can be determined from (\ref{37}) immediately.
Since $h_*\in(\sqrt{3},2)$ in Lemma 3.1, we have
\begin{equation}\label{39}\beta\geqslant\frac{h_*+1}{h_*-1}\alpha>3\alpha.\end{equation}
Hence, noting $0<\gamma^+\leqslant\alpha$ and (\ref{39}), it follows  from (\ref{37}) that
 \begin{equation}\label{40}\frac32(\alpha+\beta)>\gamma^->\frac{3\beta-\alpha}{2}>\alpha+\beta,\end{equation}
which  implies
\begin{equation}\label{41}\gamma^++\gamma^->\alpha+\beta.\end{equation}
Therefore, the dissipation rate $D(t)>0$
and this means that for the case $\beta\geqslant\frac{h_*+1}{h_*-1}\alpha$, the constructed unique solution (\ref{14}) is physically admissible.

On the other hand, direct calculation shows that $c(\gamma^+)=\dot{s}<c(\gamma^-),$
which implies that the phase boundary speed is equal to the characteristic speed in the front-state and  subsonic
in the back-state of the phase boundary. This fact can be viewed as  the geometrical feature of the phase boundary under the assumption
  $\beta\geqslant\frac{h_*+1}{h_*-1}\alpha$, while for the case  $\beta<\frac{h_*+1}{h_*-1}\alpha$, the phase boundary  exhibits a different manner, i.e., $c(\gamma^{\pm})>\dot{s}$ and $c(\gamma^+)=c(\gamma^-)$ (cf. \cite{dk}).

\begin{Remark}Using the method given in \cite{dk1}, we also can investigate the global structure stability of the physical solution constructed in this section. The details are omitted for space limitations.
\end{Remark}
\section{Concluding remarks}
This paper concerns the propagation of impact-induced tensile waves
in a kind of phase transforming materials. For the impact velocity
of intermediate strength $V_{*}<V<V_{***}$, the biggest problem
under consideration is how to select the unique physical solution.

For the material governed by (\ref{3}), we consider the impact
problem by dividing the parameters in stress-response function into
two cases: $\beta<\frac{h_*+1}{h_*-1}\alpha$ and
$\beta\geqslant\frac{h_*+1}{h_*-1}\alpha$. In both cases, we can
establish the mixed  centered rarefaction wave and phase boundary
solution (\ref{14}). It is pointed out the geometric features of the
two phase boundaries are different. The mixed-type solution in the
former case indeed satisfies the dissipation free kinetics and the
latter one corresponds to the maximally dissipative kinetics.
Precisely speaking, direct computation shows that this kinetic
relation is given by
\begin{equation*}\label{4.1}f=\frac94E\left(\frac{\beta-\alpha}{2}\right)^4
\left[1+\left(\frac{\dot{s}}{c_2}\right)^2\right]^2.\end{equation*}
 Most recently, Knowles \cite{jk1} studies the
maximally dissipation kinetics for general nonlinear elastic bar and
points out it depends strongly on the qualitative nature of the
stress-strain relation.

Finally, we would like to mention that for the intermediate impact,
under the maximally dissipative kinetics hypothesis,  the solution
containing a centered rarefaction wave and a phase boundary can be
constructed for all values of $\alpha, \beta$, see the curves
O$'$B$'$C$'$ in Fig. 4 and Fig. 5; while under the dissipation free
kinetics, construction of this mixed-type solution is not obvious
except the case $\beta<\frac{h_*+1}{h_*-1}\alpha$ as discussed in
the last section.

\vskip 10mm\noindent{\Large {\bf Acknowledgements.}} The author
would like to thank Professor Hui-Hui Dai and De-Xing Kong for
valuable discussions. This work was supported by the Foundation for
University's Excellent Youth Scholars (Grand Nos. 2009SQRZ025ZD,
2010SQRL025) and the University's Natural Science Foundation from
Anhui Province (Grand No. KJ2010A130).

\end{document}